\documentclass[10pt,letterpaper,twocolumn]{article} %% two column, final layout

\usepackage{ol2}
\usepackage[draft]{hyperref}
\usepackage{amsmath}

\begin{document}

\twocolumn[ %% activate for two-column option

\title{Jaynes-Cummings photonic superlattices}

%% For REVTeX it is possible to automate superscript and e-mail callouts with the superscriptaddress option; see REVTeX4 documentation.

\author{Stefano Longhi}

\address{Dipartimento di Fisica, Politecnico di Milano, Piazza L. da Vinci 32, I-20133 Milano, Italy}

\begin{abstract}
A classical realization of the  Jaynes-Cummings (JC) model,
describing the interaction of a two-level atom with a quantized
cavity mode, is proposed based on light transport in engineered
waveguide superlattices.  The optical setting enables to visualize in Fock space dynamical regimes not yet accessible in quantum systems, providing new physical insights into the deep strong coupling regime  of the JC model. In particular,  
   bouncing of photon number wave packets in Hilbert space and
revivals of populations are explained as generalized Bloch oscillations in an inhomogeneous tight-binding lattice.
\end{abstract}


%230.7370 Waveguides
%020.5580   Quantum electrodynamics
%000.1600
%Classical and quantum physics

 ] %% activate for two-column option

\noindent Transport of discretized light in photonic lattices has
received a great attention in recent years, suggesting the
possibility to realize novel functionalized optical materials \cite{R1,R2}.
Photonic lattices also provide a beautiful laboratory system to
realize classical analogues of a wealth of quantum phenomena ranging from solid-state physics
\cite{R1,R2,R3,B02,B03,Z1,Z2,AL,DL1,DL2} to relativistic quantum
mechanics \cite{rel1,rel2,rel3,rel4,rel5}.  In this Letter it is
shown that photonic lattices can provide classical realizations of
cavity quantum electrodynamics (QED) systems as well, offering the possibility to access and visualize dynamical regimes not yet accessible in QED systems. Here a
photonic realization of the renowned Jaynes-Cummings (JC) model,
describing the interaction of a two-level atom with a quantized mode
of a cavity \cite{JC1,JC2}, is proposed.  The 
JC model has seen a renewed interest since the advent of circuit QED, where
on-chip superconducting qubits and oscillators play the roles of
two-level atoms and cavities. In such systems,  the possibility to
explore novel dynamical regimes has been suggested, and the first
experimental observations of breakdown of the rotating wave
approximation (RWA) have been reported \cite{Q4,Q1,Q2,Q3}. Recently,
the deep strong coupling (DSC) regime of the JC model has been
theoretically investigated \cite{Q2}, and bouncing of photon number
wavepackets in Hilbert space, leading to collapse and revivals of
the initial populations, have been predicted. The DSC regime,
however, is still far from being accessible in semiconductor or
superconducting circuit QED systems. The classical realization of
the JC model proposed in this Letter provides a testbed to mimic
in the lab  {\em all} the dynamical regimes of the JC Hamiltonian directly in Hilbert space.
It also gives a new physical view onto the revival dynamics of the DSC: it can be explained as a kind of
generalized Bloch oscillations of the wave function in an inhomogeneous tight-binding lattice. 
\par The JC Hamiltonian describing a qubit
interacting with a bosonic mode [Fig.1(a)]
 reads \cite{JC2}
\begin{equation}
\hat{H}=\frac{\hbar \omega_0}{2} \hat{\sigma}_z+ \hbar \omega
\hat{a}^{\dag} \hat{a}+ \hbar g (\hat{\sigma}_+
+\hat{\sigma}_-)(\hat{a}+\hat{a}^{\dag}), \
\end{equation}
where $\hat{a}$ and $\hat{a}^{\dag}$ are the annihilation and
creation operators of the quantized oscillator with frequency
$\omega$, $\hat{\sigma}_z=|e \rangle \langle e|-|g \rangle \langle
g|$, $\hat{\sigma}_{+}=|e \rangle \langle g|$ and
$\hat{\sigma}_{-}=|g \rangle \langle e|$ are Pauli operators
associated to the qubit with ground state $|g\rangle$, excited state
$|e \rangle$, and transition frequency $\omega_0$, and $g$ is the
coupling strength. The usual JC model under the RWA is obtained in
the weak-coupling and near-resonant limits ($|\omega-\omega_0| \ll
\omega_0$, $ \omega \gg g$), and corresponds to neglect the
anti-resonant interaction terms $\hat{\sigma}_+ \hat{a}^{\dag}$ and
$\hat{\sigma}_- \hat{a}$ in the Hamiltonian (1). On the other hand,
the DSC limit corresponds to a coupling $g$ of the order or larger
that the oscillator frequency $\omega$, with no particular relation
between $\omega$ and $\omega_0$ \cite{Q2}. The main idea underlying
the proposed photonic realization of the JC model is that the
temporal evolution of the quantum system in Hilbert space (spanned
by the states $ |g \rangle |n \rangle$ and $|e \rangle |n \rangle$
describing $n$ quanta of the field with the atom in the ground or in
the excited state)
 can be mapped into the spatial propagation of light
waves in a semi-infinite curved binary array of waveguides with
non-uniform coupling rates. In fact, let us expand the vector state
of the system as $|\psi(t) \rangle=\sum_{n=0}^{\infty} [a_n(t)
|e\rangle |n \rangle +b_n(t) |g \rangle |n \rangle]$ and let us
introduce the amplitudes $c_n(t)$ and $f_n(t)$, defined by
$c_n(t)=a_n \; , f_n(t)=b_n$  for $n$ even, and $c_n(t)=b_n \; ,
f_n(t)=a_n$ for $n$ odd. Then, the  temporal evolution of the
amplitudes $c_n$ and $f_n$, as obtained  by the Schr\"{o}dinger
equation $i \hbar
\partial_t | \psi(t) \rangle=\hat{H} |\psi(t) \rangle$, read
\begin{eqnarray}
i \frac{dc_n}{dt} & = & \kappa_n c_{n+1}+\kappa_{n-1}c_{n-1}
+(-1)^n \frac{\omega_0}{2}c_n+n \omega c_n \;\;\;\;\; \\
i \frac{d f_n}{dt} & = & \kappa_{n} f_{n+1}+\kappa_{n-1} f_{n-1}
-(-1)^n \frac{\omega_0}{2}f_n+n \omega f_n  \;\;\;\;\;\;
\end{eqnarray}
($n=0,1,2,3,...$), where we have set
$\kappa_n=g \sqrt{n+1}$.
\begin{figure}[htb]
\centerline{\includegraphics[width=8.2cm]{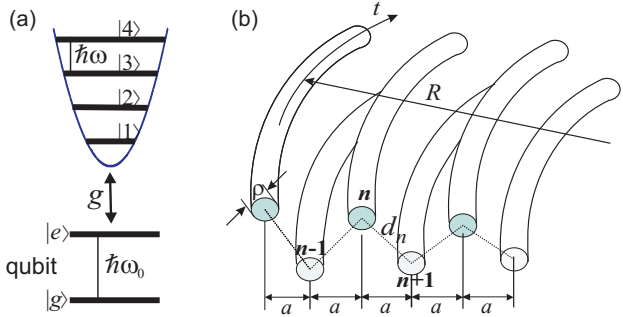}} \caption{
(Color online) (a) Schematic of a qubit interacting with a single
bosonic mode. (b) Photonic realization of the JC model in Hilbert
space based on a semi-infinite curved binary waveguide array.}
\end{figure}
In the optical context, Eqs.(2) and (3) describe light transport in
two uncoupled semi-infinite binary photonic lattices in the
tight-binding approximation with a superimposed transverse index
gradient and with non-uniform coupling constant $\kappa_n$ between
adjacent waveguides $n$ and $(n+1)$ (see, for
instance, \cite{Z2}). In the optical analogue, the temporal variable
$t$ plays the role of the spatial propagation distance, the
transition frequency $\omega_0$ of the qubit corresponds to the
propagation constant mismatch of alternating waveguides of the
superlattice, and the frequency $\omega$ of the quantized mode to
the transverse index gradient. The superlattices (2) and (3) are
decoupled and basically correspond to the two uncoupled parity
chains of the Hilbert space discussed in Ref.\cite{Q2}. Since the
superlattices (2) and (3) are obtained each other by reversing the
sign
 of $\omega_0$, the analysis can be limited to consider the dynamics of either one of the two lattices. Their optical realization
can be obtained using a semi-infinite chain of circularly-curved
waveguides with alternating propagation constants and with
engineered distances $d_n$ in the geometrical setting depicted in
Fig.1(b) \cite{LonghiPRB2009}. The index gradient $\omega$
 is given by $\omega=2 \pi n_s a/(R
\lambda)$, where $n_s$ is the refractive index of the substrate,
$\lambda$ is the wavelength of light, $R$ the bending radius of curvature, and $a$ is their horizontal spacing
\cite{LonghiPRB2009}. It should be noted that in the limiting case
$\omega=\omega_0=0$ the lattice reduces to the one recently proposed
in Ref.\cite{Szameit} as a classical analogs to quantum coherent and
displaced Fock states. \par The usual RWA of the JC model
corresponds to strongly detuned waveguides, $g/ \omega_0 \rightarrow
0$, with the resonance condition $\omega \simeq \omega_0$, which
decouple light dynamics of the array in pairs of waveguides (i.e. in
a sequence of optical directional couplers). In fact, if we consider
for example the lattice (2), after setting $c_n=\theta_n \exp[-i
(-1)^n \omega_0 t/2-i n \omega t]$, in the RWA approximation from
Eq.(2) a set of decoupled equations are obtained for the
slowly-varying amplitudes $ \theta_n$ and $\theta_{n+1}$ describing
the occupation amplitudes of states $|e \rangle |n \rangle$ and $|g
\rangle |n+1 \rangle$, respectively
\begin{equation}
i \frac{d \theta_n}{dt} =  \kappa_{n} \exp(-i \Delta t) \theta_{n+1}
\; , \; i \frac{d \theta_{n+1}}{dt} =  \kappa_{n} \exp(i \Delta t)
\theta_n \nonumber
\end{equation}
\begin{figure}[htb]
\centerline{\includegraphics[width=8.2cm]{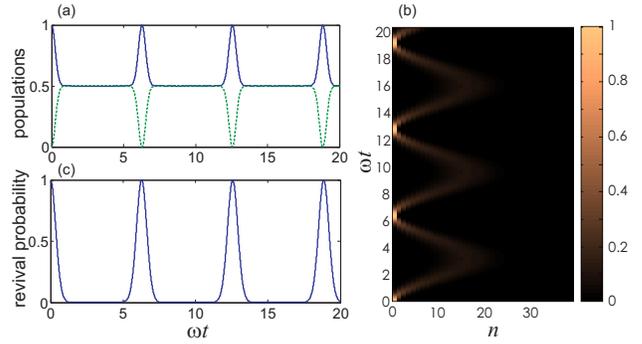}} \caption{
(Color online) (a) Behavior of populations $P_g$ (solid curve) and
$P_e$ (dashed curve) of ground ($|g\rangle$) and excited
($|e\rangle$) qubit states versus normalized time $\omega t$ for the
JC Hamiltonian (1) in the DSC regime for initial state $|\psi(0)
\rangle=|g \rangle |0 \rangle$. Parameter values are $g/ \omega=2$
and $\omega_0/ \omega =0$. The corresponding behavior of photon
statistics $P(n,t)=|f_n(t)|^2$ and revival probability
$P_{rev}=|\langle \psi(t)|\psi(0) \rangle|^2$ are shown in (b) and
(c).}
\end{figure}
where $\Delta=\omega-\omega_0$ and the integer $n$ is taken to be
even. Hence, Rabi-like oscillations of the optical power between the
two waveguides $n$ and $(n+1)$ with spatial frequency
$\Omega_n=\sqrt{g^2(n+1)+(\Delta /2)^2}$ is obtained, which is the
analog of the QED Rabi frequency \cite{JC2}. \par A more exotic
dynamics is observed in the DSC regime of the JC model \cite{Q2}. In
the limit of degenerate qubit levels, i.e. for $\omega_0 / \omega
\rightarrow 0$, the energies and eigenstates of the JC Hamiltonian
can be determined analytically, and the dynamics of populations and
photon numbers is strictly periodic with period $T=2 \pi/ \omega$
\cite{R2}. As an example, in Fig.2 it is shown the dynamical
evolution of populations in the two qubit levels, $P_g(t)$ and
$P_e(t)=1-P_g(t)$, and of the photon number distribution $P(n,t)$ as
obtained from numerical analysis of the JC Hamiltonian (1) when the
system is initially prepared in the pure state $|\psi(t=0)
\rangle=|g \rangle |0 \rangle$. In the figure, the evolution of the
return (revival) probability to the initial state,
$P_{rev}(t)=|\langle \psi(t)| \psi(0) \rangle |^2$, is also
depicted. The behavior of $P_{rev}(t)$ clearly shows the existence
of collapse and (exact) revivals, which have been explained in
Ref.\cite{Q2} as due to the periodic bouncing of photon number
wavepackets along one of the two parity chains in the Hilbert space,
as shown in Fig.2(b). Such a behavior is deeply distinct from the
usual collapse-revival dynamics of the JC model in the RWA, which
requires large initial coherent states \cite{JC2}. In our optical
realization of the JC model, the evolution of the light beam
intensity along the array of Fig.1(b) when the boundary ($n=0$)
waveguide is excited at the $t=0$ input plane exactly reproduces the
evolution of the photon number distribution shown in Fig.2(b). In
fact, the populations $P_g(t)$ and $P_e(t)$, the photon number
distribution $P(n,t)$ and revival probability $P_{rev}(t)$  are
related to the occupation amplitudes $f_n(t)$ of light beam in the
waveguides of the lattice according to $P_g(t) =
 \sum_{n=0}^{\infty}|f_{2n}(t)|^2$,
$P_e(t)=\sum_{n=0}^{\infty}|f_{2n+1}(t)|^2$, $P(n,t) = |f_n(t)|^2$,
and $P_{rev}(t)=|f_0(t)|^2$.
It should be noted that the  proposed optical analog simulator gives a new physical 
view of the revival dynamics of the JC model in the DSC regime not noticed in previous work \cite{Q2}. Indeed, 
the periodic bouncing of the photon wave packet in Hilbert space shown in Fig.2(b) is the signature of the
{\it self-imaging} property of the waveguide lattice, and the observed oscillatory beam path can be ultimately explained in terms of 
a kind of {\it generalized Bloch oscillations} in an inhomogeneous tight-binding lattice under a dc bias. 
In fact, in the limit $\omega_0 / \omega \rightarrow 0$ the lattice
described by Eq.(3) belongs to a class of exactly-solvable lattice
models which shows an equally-spaced Wannier-Stark ladder spectrum
with energies $E_l=l \omega-g^2/ \omega $ ($l=0,1,2,3,...$), as
shown in Ref. \cite{Longhi2010}. Note that, as compared to most common optical Bloch oscillations previously investigated in {\it homogeneous} lattices \cite{R1,B02,B03}, in our case 
the waveguide lattice turns out to be {\it truncated} and {\it inhomogeneous}. However, the main oscillatory dynamics of the optical beam can be 
still explained from a semiclassical analysis of the inhomogeneous lattice Hamiltonian as a result of beam acceleration induced by the dc bias field [the last term on the right hand side of Eq.(3)] and Bragg scattering in the lattice. Moreover, since for $\omega_0 / \omega \rightarrow 0$ the JC model is realized by a {\it single band} tight-binding lattice, Zener tunneling \cite{Z1,Z2} or other phenomena, like interband Rabi oscillations of Bloch modes, do not occur in our case. \par

Typical design parameters of the
lattice that realizes the dynamics of Fig.2 can be obtained for
femtosecond laser written waveguides in fused silica \cite{R2}
following the lines detailed in Ref.\cite{LonghiPRB2009}. For
instance, if we assume circular waveguides with core diameter $
\rho=5 \; \mu$m and index change $\Delta n=0.002$  (as in Ref.
\cite{LonghiPRB2009}), for an axis bending radius $R=60$ cm, an
horizontal waveguide spacing $a=6 \; \mu$m and a bulk refractive
index $n_s=1.45$, at the wavelength $\lambda=633$ nm the period of
Bloch oscillations (i.e. of the revival) shown in Fig.2(b) turns out
to be $T= 2 \pi / \omega \simeq 4.37$ cm. The coupling constants are
given by $\kappa_n = g \sqrt{n+1}$ with $g=2 \omega \simeq 0.288 \;
{\rm mm}^{-1}$. The distances $d_0$, $d_1$, $d_2$, ... between
adjacent waveguides [see Fig.1(b)] that realize such non-uniform
coupling constants are calculated by the approximate relation
\cite{LonghiPRB2009} $d_n \simeq (1/ \gamma) {\rm ln}(A/ \kappa_n)$
with $A \simeq 24.6 \; {\rm mm}^{-1}$ and $\gamma \simeq 0.466 \;
{\rm \mu m}^{-1}$, i.e. $d_0 \simeq 9.54 \; \mu$m, $d_1 \simeq 8.80
\; \mu$m, $d_2 \simeq 8.37 \; \mu$m, etc. Note that a
 rather limited number of waveguides, less
than $\sim 25$ [see Fig.2(b)], is enough to visualize the dynamics
in the Hilbert space. Finally, it should be noted that a nonvanishing value of $\omega_0 / \omega$ makes the
revivals only approximate, as shown in Fig.3.  In this case, the lattice is a {\it binary} one and the Wannier Stark states are no more 
equally space. However, even though the wave packet revival
is incomplete, photon bouncing is clearly observable over a few cycles.
\par
\begin{figure}[htb]
\centerline{\includegraphics[width=8.2cm]{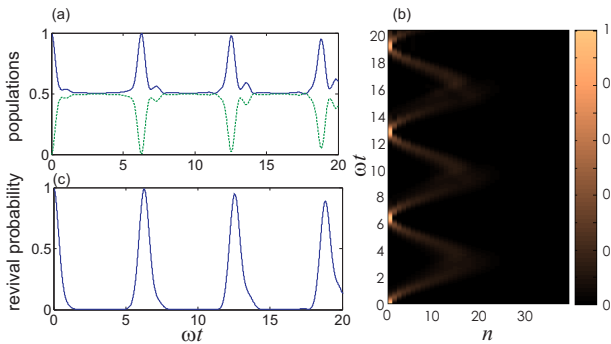}} \caption{
(Color online) Same as Fig.2, but for $g/ \omega=2$ and $\omega_0/
\omega =0.3$.}
\end{figure}
In conclusion, an optical realization of the  JC model
of QED, based on light transport in waveguide superlattices, has
been proposed. Our optical setting enables to simulate the physics
of atom-field interaction in the deep strong coupling regime
\cite{Q2}, which is currently not accessible in cavity or circuit
QED systems.
\par
Work supported by the italian MIUR (Grant No. PRIN-2008-YCAAK).

\newpage

\footnotesize {\bf References with full titles}\\
\\
 1. D. Christodoulides, F. Lederer and Y. Silberberg,
"Discretizing light behavious in linear and nonlinear waveguide
lattices", Nature
{\bf 424}, 817 (2003).\\
 2. A. Szameit and S. Nolte, "Discrete optics in
femtosecond-laser-written photonic structures," J. Phys. B {\bf 43},
163001 (2010).\\
 3. S. Longhi, "Quantum-optical analogies using photonic
structures", Laser and Photon. Rev. {\bf 3}, 243 (2009).\\
 4. T. Pertsch, P. Dannberg, W. Elflein, A. Br\"{a}uer, and F.
Lederer, "Optical Bloch Oscillations in Temperature Tuned Waveguide
Arrays," Phys. Rev. Lett. {\bf 83}, 4752 (1999).\\
 5. R. Morandotti, U. Peschel, J. Aitchinson, H. Eisenberg, and Y.
Silberberg, "Experimental observation of linear and nonlinear
optical bloch oscillations," Phys. Rev. Lett. {\bf 83}, 4756
(1999).\\
 6. H. Trompeter, T. Pertsch, F. Lederer, D. Michaelis, U. Streppel,
A. Br\"{a}uer, and U. Peschel, "Visual Observation of Zener
Tunneling," Phys. Rev. Lett. {\bf 96}, 023901 (2006).\\
 7. F. Dreisow, A. Szameit, M. Heinrich, T. Pertsch, S. Nolte, A.
T\"{u}nnermann, and S. Longhi, "Bloch-Zener Oscillations in Binary
Superlattices", Phys. Rev. Lett. {\bf 102}, 076802 (2009).\\
 8. T. Schwartz, G. Bartal, S. Fishman and M. Segev, "Transport and
Anderson Localization in disordered two-dimensional Photonic
Lattices", Nature {\bf 446}, 52 (2007).\\
 9. S. Longhi, M. Marangoni, M. Lobino, R. Ramponi, P. Laporta, E.
Cianci, and V. Foglietti, "Observation of dynamic localization in
periodically curved waveguide arrays", Phys. Rev. Lett. {\bf 96},
243901 (2006).\\
 10. A. Szameit, I.L. Garanovich, M. Heinrich, A.A. Sukhorukov, F.
Dreisow, T. Pertsch, S. Nolte, A. T\"{u}nnermann, and Y.S. Kivshar,
"Polychromatic dynamic localization in curved photonic lattices",
Nature Physics {\bf 5}, 271 (2009).\\
 11. X. Zhang, "Observing Zitterbewegung for Photons near the Dirac
Point of a Two-Dimensional Photonic Crystal", Phys. Rev. Lett. {\bf
100}, 113903 (2008).\\
 12. O. Bahat-Treidel, O. Peleg, M. Grobman, N. Shapira, T.
Pereg-Barnea, and M. Segev, "Klein Tunneling in Deformed Honeycomb
Lattices", Phys. Rev. Lett. {\bf 104}, 063901 (2010).\\
 13. S. Longhi, "Photonic analog of Zitterbewegung in binary
waveguide arrays", Opt. Lett. {\bf 35}, 235 (2010).\\
 14. S. Longhi, "Klein tunneling in binary photonic superlattices",
Phys. Rev. B {\bf 81}, 075102 (2010).\\
 15. F. Dreisow, M. Heinrich, R. Keil, A. T\"{u}nnermann, S. Nolte,
S. Longhi, and A. Szameit, "Classical Simulation of Relativistic
Zitterbewegung in Photonic Lattices", Phys. Rev. Lett. {\bf 105},
143902 (2010).\\
 16. E.T. Jaynes and F.W. Cummings, "Comparison of quantum and
semiclassical radiation theories with application to the beam
maser", Proc. IEEE {\bf 51}, 89  (1963).\\
 17. C.C. Gerry and P.L. Knight, {\it Introductory Quantum Optics}
(Cambridge, Cambridge University Press, 2004).\\
 18. A. G\"{u}nter, A. A. Anappara, J. Hees, A. Sell, G. Biasiol, L.
Sorba, S. De Liberato, C. Ciuti, A. Tredicucci, A. Leitenstorfer,
and R. Huber, "Sub-cycle switch-on of ultrastrong light-matter
interaction", Nature {\bf 458}, 178 (2009).\\
 19. T. Niemczyk, F. Deppe, H. Huebl, E. P. Menzel, F. Hocke, M. J.
Schwarz, J. J. Garcia-Ripoll, D. Zueco, T. H\"{u}mmer, E. Solano, A.
Marx, and R. Gross, "Beyond the Jaynes-Cummings model: circuit QED
in the ultrastrong coupling regime", Nature Physics {\bf 6}, 772
(2010).\\
 20. J. Casanova, G. Romero, I. Lizuain, J. J. García-Ripoll, and E.
Solano, "Deep Strong Coupling Regime of the Jaynes-Cummings Model",
Phys. Rev. Lett. {\bf 105}, 263603 (2010).\\
 21. P. Forn-Díaz, J. Lisenfeld, D. Marcos, J. J. García-Ripoll, E.
Solano, C. J. P. M. Harmans, and J. E. Mooij, "Observation of the
Bloch-Siegert Shift in a Qubit-Oscillator System in the Ultrastrong
Coupling Regime", Phys. Rev. Lett. {\bf 105}, 237001 (2010).\\
 22. S. Longhi, "Bloch oscillations and Wannier-Stark localization
in a tight-binding lattice with increasing intersite coupling ",
Phys. Rev. B {\bf 80}, 033106 (2009).\\
 23. A. Perez-Leija, H. Moya-Cessa, A. Szameit, and D.N.
Christodoulides, "Glauber-Fock photonic lattices", Opt. Lett. {\bf
35}, 2409 (2010).\\
 24. S. Longhi, "Periodic wave packet reconstruction in truncated
tight-binding lattices", Phys. Rev. B {\bf 82}, 041106 (2010).

\end{document}